\documentstyle[12pt]{article}
\begin{document}

\begin{center}

{\Huge Scaling in a Multispecies 

\vspace{0.2 cm}

Network Model Ecosystem}

\vspace{0.6 in}

{\large Ricard V. Sol\'e$^ {1,2}$, David Alonso$^{1,3}$ and Alan McKane$^4$}

\vspace{0.6 in}

(1) Complex Systems Research Group

Department of Physics, FEN, Universitat Polit\`ecnica de Catalunya

Campus Nord B4, 08034 Barcelona, Spain

\vspace{0.3 cm}

(2) Santa Fe Institute, 1399 Hyde Park Road, New Mexico 87501, USA

\vspace{0.3 cm}

(3) Department of Ecology, Facultat de Biologia, Universitat de Barcelona

Diagonal 645, 08045 Barcelona, Spain

\vspace{0.3 cm}

(4) Department of Theoretical Physics, University of Manchester

Manchester M13 9PL, UK

\vspace{1 cm}

\begin{abstract}

A new model ecosystem consisting of many interacting species is introduced. 
The species are connected through a random matrix with a given connectivity 
$C_{\pi}$. It is shown that the system is organized close to a boundary of 
marginal stability in such a way that fluctuations follow power law
distributions both in species abundance and their lifetimes for some 
slow-driving (immigration) regime. The connectivity and the number of species
are linked through a scaling relation  which is the one observed in real 
ecosystems. These results suggest that the basic macroscopic features of 
real, species-rich ecologies might be linked with a critical state.
A natural link between lognormal and power law distributions of species 
abundances is suggested.

\end{abstract}

\vspace{0.5 cm}

{\large Submitted to Physical Review Letters}

\vspace{0.5 cm}

{PACS number(s): 87.10.+e, 05.40.+j, 05.45.+b}

\end{center}

\baselineskip=7 mm

\newpage

The universal properties of complex biosystems have attracted the attention
of physicists over the last decade or so. The standard approach to complex 
ecosystems is based on the classical Lotka-Volterra (LV) 
$S$-species model [1]:
$${d N_i \over dt} = N_i \Biggl ( \epsilon_i - 
\sum_{j=1}^S \alpha_{ij} N_j(t) \Biggl )\ , \eqno (1)$$
where $\{N_i\}; (i=1,...,S)$ is the population size of each species. 
Here $\epsilon_i$ and $\alpha_{ij}$ are constants that introduce feedback
loops and interactions among different species. The matrix 
${\bf A} = (\alpha_{ij})$ describes the interaction graph [1] and 
even a small degree of asymmetry leads to very complicated dynamics, although
several generic properties have been identified. In this
respect, recent studies on a related class of replicator equations,
where an initial random graph evolves towards a highly non-random
network, revealed remarkable self-organized features [2].

A classical result on randomly wired ecosystems [3]
shows that a critical limit to the number of species $S$
exists for a given connectivity $C$. Here $C$ is the fraction of non-zero
matrix elements. This limit is sharp (a phase transition):
below the critical value the system is stable but it becomes unstable 
otherwise. The basic inequality is that the community will be stable if
$C \le \sigma/S$, where $\sigma$ is a given constant. 
Here the equality defines the boundaries between the stable
and the unstable domains. Field studies show 
that, in general, $C = \sigma S^{-1+\epsilon}$ with $\epsilon 
\in [0,1/2]$ [4].

More recently some authors have suggested that complex 
biosystems could be self-organized into a critical state [5,6]. Most of these
approaches based on self-organized criticality (SOC), such as the 
Bak-Sneppen model [5], involve a time scale that is assumed to be very large. 
But the species properties of a given ecosystem do not change
appreciably on a smaller time scale [7] and 
several field observations suggest that real
ecosystems display some features characteristic of SOC states.
In particular: (i) the analysis of time series 
[8] shows that the largest Lyapunov exponent is typically close to
zero; (ii) studies on colonization in islands show that after a critical
number of species is reached, extinction events are triggered, following a
power law distribution [9]; (iii) the lifetime distribution of species
(defined in terms of local extinction in given areas) is also 
a power law $N(T) \approx T^{-\theta}$ with $\theta=1.12$ [9] to $\theta=1.6$ [10] and
(iv) in species-rich ecosystems, the number of species $N_s(n)$
with $n$ individuals is also a power law $N_s(n) \approx n^{-\gamma}$,
where $\gamma$ is close to one, although log-normal distributions with long 
tails are also well known [11]. But none of these observations (together with
the $C-S$ law) have been explained within a single theoretical framework.

\begin{figure}[htb]
\vspace{7.5 cm}
\includegraphics{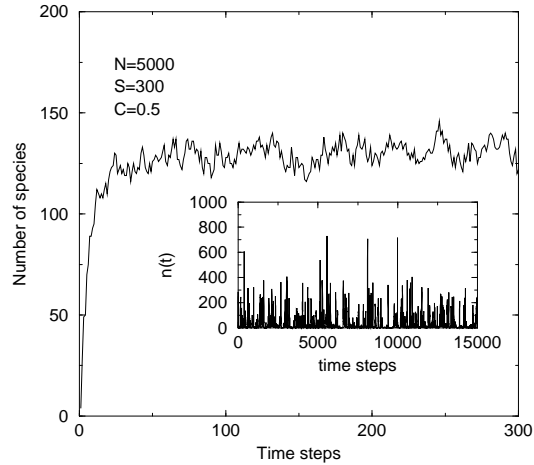}
\caption{An example of the time evolution of a single species (inset) from a 
population of $N=5000$ individuals and $S=300$ species in the pool. Here 
$C=0.5$ and $\mu=5 \times 10^{-3}$. We can see a wide spectrum of 
fluctuations in $n(t)$. Main figure: the number of species present in the 
simulated ecosystem increases up to an average value $<S>\approx 125$ imposed
by the matrix connectivity. }
\end{figure}

In this letter we present a new model which can be used as a simple 
alternative to the LV formulation and follows the simple (but somewhat 
different) approach of other studies on random graphs [2]. The model involves a
 set $\Theta_s$ of $S$ (possible) species and 
a population size $N$. Individuals of different species interact
through a matrix $\bf \Omega$, with a pre-defined connectivity $C_{\pi} \in 
[0,1]$. The $\Omega_{ij}$ are randomly chosen from a uniform
distribution $\rho(x)$.  Each time step two individuals are taken at random 
belonging, say, to species $i$ and
$j$ respectively. Then if $\Omega_{ij}>\Omega_{ji}$ the individual belonging
to $j$ is replaced, with probability $1-\mu$, by a new individual of species 
$i$. If $\Omega_{ij}=\Omega_{ji}=0$ then nothing happens. This rule is 
repeated $N$ times and these $N$ updates define our time step. Additionally, 
with probability $\mu$, any individual can be replaced by another individual
of {\it any} of the species from $\Theta_s$. These rules makes our model 
close to equation (1) when $\alpha_{ij}>0$ ($S$-species competition model [1]).
So two basic reactions are allowed: (i) interaction through $\bf \Omega$ and
(ii) random replacement by a new species from the $\Theta_s$ pool.
The second rule introduces immigration (at a rate $\mu$): our driving 
force [12,13].

An example of the temporal dynamics of this model is shown in figure 1,
 where we can see a wide spectrum of 
fluctuation sizes, not very different from those observed in some natural 
communities [1,8]. Starting from any given initial number of
species, this model self-organizes towards a subset of species (with average 
$<S>$). The specific subset itself changes in time. Let $C^*$ be the 
probability of having two species connected. Then $C^*$ and $<S>$ are linked by a power law 
$<S>=k(C^*)^{-1+\epsilon(\mu)}$. Once this $<S>$ is reached, complex 
fluctuations can arise and under some conditions (see below) 
power laws are observable. The analysis of the population fluctuations 
typically shows, for small driving $\mu$ and large enough $C_{\pi}$, a power 
law $P(n)\approx n^{-\gamma} f_{\mu}(n)$. Here $\gamma=1$ when 
$N \gg S \gg 1$ and $f_{\mu}(n)$ an exponential cut-off. The lifetime 
distribution of species is also a power law: $N(T)\approx T^{-\theta}$ and 
for $N \gg S$, $\theta \in (1,3/2)$. The $\theta=3/2$ limit
corresponds to low-level connectivities (i.e. random-walk-like behavior) and
$\theta=1$ to the situation were interactions dominate over immigration. It
is remarkable that these limits in the lifetimes correspond to those reported
from field studies [9,10].

\begin{figure}[htb]
\vspace{8.5 cm}
\includegraphics{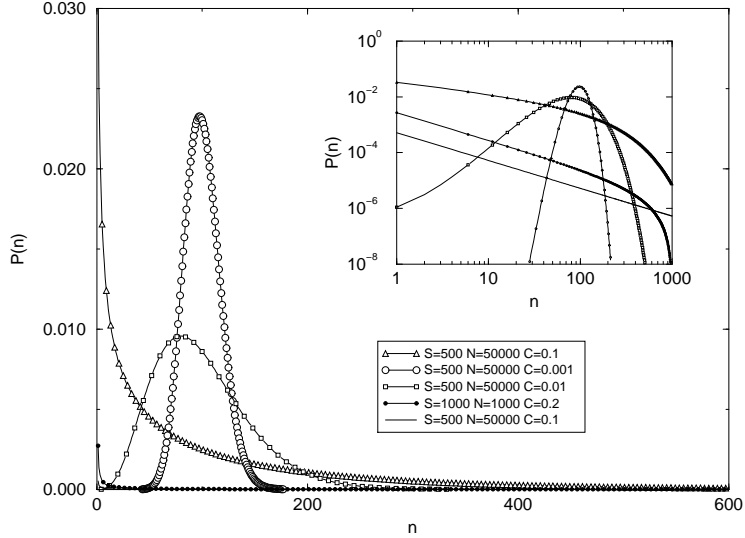}
\caption{Stationary probability distribution $P_s(n)$ obtained from (4)-(6), here 
$\mu=10^{-3}$. For different values of $S/N$ and $C$ we get different 
stationary distributions. For $S=500, N=50000$ and different connectivities, 
we get Gaussian $(C=0.001$), lognormal $(C=0.01$) or power laws. A power law 
with a perfect scaling is shown for $\mu=10^{-6}$ and 
$C=0.1$ (here $P(n)\approx n^{-1}$). We also show a power law with exponential
bending (black squares). }
\end{figure}

The limiting case for $\mu \rightarrow 0$ shows the exact hyperbolic relation
and can be derived from a mean-field approach. It is not difficult to see
that the number of species will change in time following:
$$ {d S  \over dt} = \mu N - C^* S\ , \eqno(2)$$
where the first term on the right-hand side stands for the
continuous introduction of new species at a rate $\mu$ and the second
for the instability derived from interactions. The previous
equation gives an exponential approach to criticality
$S(t) = [\mu N - (\mu N - C^* S(0) ) e^{-C^* t}]/C^*$ which
leads at the steady state to 
$$ S^* = {\mu N \over C^*} \propto {1 \over C^*}\ . \eqno(3)$$ 

A  mean field theory of species abundance can be derived. Let us start
with the master equation for this model. If $P(n,t)$ is the probability 
(for any species) of having $n$ individuals at 
time t,  the one-step process is described by [14]:
$${dP(n,t) \over  dt} = r_{n+1} P(n+1,t) + g_{n-1}
P(n-1,t) - (r_n + g_n) P(n,t)\ , \eqno(4)$$
where the one-step transition rates are:
$$r_n \equiv W(n-1 \vert n)
=   C^* (1-\mu){n \over N} {N-n \over N-1} + {\mu \over S} (S-1){n\over N}\ ,
\eqno(5)$$  
$$g_n \equiv W(n+1 \vert n) = C^* (1-\mu){n \over N} {N-n \over
N-1} + {\mu \over S} (1-{n\over N})\ .  \eqno(6)$$
From (5) and (6) we see that $r_0=0$ and 
$g_N=0$ and we define $g_{-1}=0$ and $r_{N+1}=0$. The stationary distribution 
$P_s(n)$ is obtained from $dP(n)/dt=0$ using standard methods [14] 
namely by writing 
$$ P_s(n) = {g_{n-1}\, g_{n-2} ...\, g_0 \over r_n\, r_{n-1} ...\, r_1} P_s(0)
\ , \eqno(7)$$
and using the normalization condition 
$\sum_{n=0}^N P_s(n) = P_s(0) + \sum_{n>0} P_s(n) =1$, to show that
$$ (P_s(0))^{-1} = \sum_{n=0}^N {N \choose n} (-1)^n 
{\Gamma (n + \lambda^*) \over \Gamma(\lambda^*)} {\Gamma (1-\nu^*)
\over \Gamma(n + 1 - \nu^*)}\ . \eqno(8)$$
Here $\lambda^*=\mu^*(N-1)$ and $\nu^*=N+\mu^*(N-1)(S-1)$, where
$\mu^*=\mu/[(1-\mu)SC^{*}]$. This sum takes the form of a Jacobi polynomial 
$P_N^{(\alpha, \beta)}(x)$ [15], with $\alpha =-\nu^*$,  
$\beta =\lambda^*+\nu^*-(N+1)$ and $x = -1$, which can itself be expressed in 
terms of gamma functions for this value of $x$. Therefore using 
$\beta(p,q)=\Gamma(p) \Gamma(q) / \Gamma(p+q)$, the stationary, normalized 
solution can be written in the form [16]
$$ P_s(n) = {N \choose n} {\beta(n+\lambda^*, \nu^*-n) \over \beta(\lambda^*,
\nu^*-N) }\ .  \eqno(9)$$
As the connectivity (or the immigration rate) increases, the distributions 
changes from Gaussian to lognormal and to power laws (fig. 2). This is 
expected since very small connectivities will make interactions irrelevant, 
with fluctuations dominated by random events. We can simplify (9) in the 
region of interest: $N$ and $S$ large, and $N \gg n_{max}$ where
$n = 1, ..., n_{max}$. If in addition $\lambda \ll 1$, which 
corresponds to  $\mu$ small: $\mu \ll SC^*/N$, we get a scaling relation:
$$P_s(n) = {\cal K} n^{-1} \exp ( - n \mu^* S)\ , \eqno(10)$$
where ${\cal K} =  (\mu^*S)^{\lambda^*}/\Gamma(\lambda^*)$  
in agreement with simulations and field data. These results suggest
a simple link between a range of statistical distributions from those 
generated by multiplicative events to those related with SOC dynamics.

\begin{figure}[htb]
\vspace{9 cm}
\includegraphics{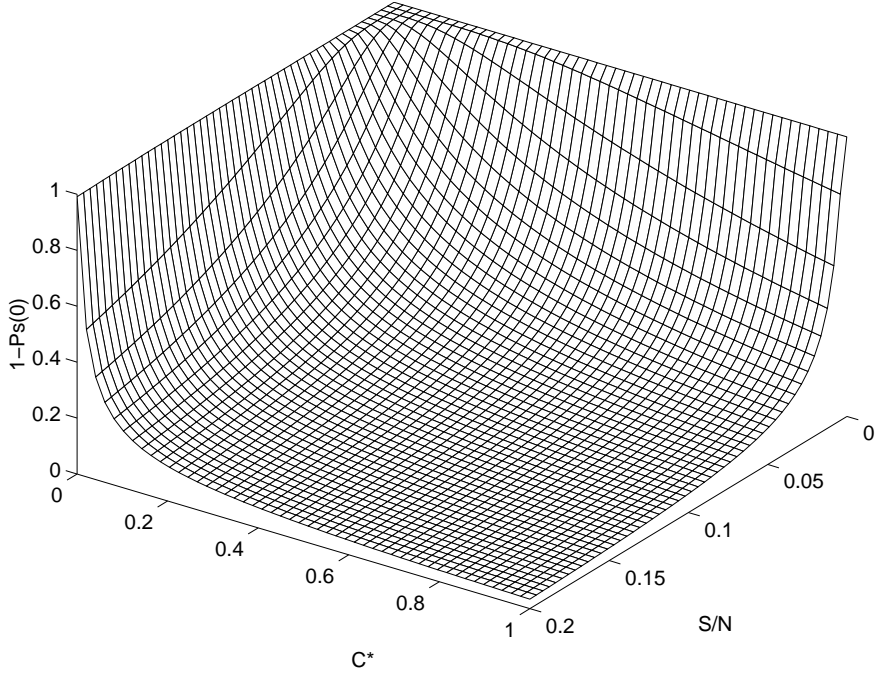}
\caption{Fraction of remaining species at steady state as a function of $S/N$ and 
connectivity ($N=50000$, $\mu=0.001$). In this parameter range, 
hyperbolic-like  behavior can be observed for high values of $S/N$.}
\end{figure}

From this solution we can compute the stationary number of species 
$<S>=(1-P(0))S$ and find the relation between this quantity and the 
probability of having two species connected, $C^*$. Under the same conditions 
that led to (10), we find that $P_s(0) = (\mu^* S)^{\lambda^*}$ [17] which
in this parameter range gives $<S>=S\mu^*N \ln(1/(\mu^*S)$. Substituting in 
the expression for $\mu^*$ gives $$<S> = {\mu N  \over (1-\mu)C^*}  
\Bigl [ \ln \Bigl ({1-\mu \over \mu} \Bigr ) + \ln C^* \Bigr ]
\approx {\cal A} (C^*)^{-1+\epsilon(\mu)}\ ,  \eqno(12)$$
where $\epsilon$ is given by $\epsilon^{-1} = \ln((1-\mu)/\mu) \approx 
\ln \mu^{-1}$ and ${\cal A}$ is a constant equal to $N\epsilon^{-1} 
\exp(-\epsilon^{-1})$. The additional conditions under which (12)
hold are $\lambda^* \ll \epsilon$ and $\vert \ln(C^*) \vert \ll \vert 
\ln(\epsilon) \vert$. The latter means that $C^*$ may not be taken too small, 
for example, if $\mu \approx 10^{-5}$ then $C^* \ge  0.1$ . With these values
the former condition is satisfied if $N/S$ is not too large. Thus
(12) holds for a wide range of parameter values. The relation 
between $<S>$ and $C^*$ is shown in figure 3 for different $S/N$ values and 
a quite large immigration rate ($\mu=10^{-3}$). In this case, the 
hyperbolic relation is only reached for large $S/N$ and for smaller $S/N$ 
values, slower decays (i. e. $<S>\approx C^{-1+\epsilon}$) are obtained [18].

In summary, a simple $S$-species network model that shows both lognormal 
(even Gaussian) distributions as well as SOC dynamics has been introduced.
Starting from random 
initial conditions, the system evolves towards a state characterized by some 
well-defined scaling laws and statistical features. High immigration, small 
$S/N$ relations or low connectivities lead to lognormal distributions. These 
are replaced by well-defined power laws as the interactions become more 
relevant. In the last case, a multiplicity of metastable states, the presence
of a threshold in the number of species $S_c=<S>$, the scale separation
between the driving when $\mu$ is small and the system response fully 
describe our system as SOC. As a consequence of critical dynamics several
well-known field observations are recovered. The interpretation of this model
provides a new framework for understanding how complex dynamics emerge in
multispecies ecosystems and suggest that some well-known time series from 
natural communities would be the result of criticality instead of 
deterministic chaotic behavior.

\vspace{0.6 cm}

{\Large Acknowledgments}

\vspace{0.2 cm}

The authors thank H. Jensen, R. Engelhardt, S. Pimm, S. Jain,  
S. Kauffman and K. Sneppen for useful comments. Special thanks to Per Bak 
for encouraging comments. This work has been supported by a grant PB97-0693 
and by the Santa Fe Institute (RVS), CIRIT 99 (DA) and EPSRC grant GR/K79307 
(AM).

\newpage

\section{References}

\begin{enumerate}

\item 
R. M. May, {\em Stability and complexity in model ecosystems}.
Princeton U. Press, Princeton (1973);  
N. S. Goel, S. C. Maitra and E. W. Montroll. Rev Mod. Phys 65, 231 (1971)

\item
S. Jain and S. Krishna, preprint adap-org/9810005, to appear in Phys. Rev. 
Lett.; see also: adap-org/9809003.

\item
R. M. May, Nature 238, 413 (1972); see also T. Hogg, B. Huberman and
J. M. McGlade, Proc. R. Soc. London B237, 43 (1989).
Specifically, if $P(S, C, \sigma)$ is the probability that the system
is stable for a given $(S, C, \sigma)$ set, where $\sigma$ is the
matrix variance, we have (for large $S$)
$P(S, C, \sigma) \rightarrow 1$ if $CS\le\sigma$ and
$P(S, C, \sigma) \rightarrow 0$ otherwise.

\item
S. Pimm, {\em The Balance of Nature}, Chicago U. Press, 1994

\item 
P. Bak, and K. Sneppen, Phys. Rev. Lett. 71, 4083 (1993) 

\item
R. V. Sol\'e and S. C. Manrubia, Phys. Rev. E54, R42-R45 (1996); 
R. Engelhardt, {\em Emergent Percolating Nets in Evolution}, Ph.D. Thesis,
University of Copenhagen, Denmark, 1998

\item
J. Maynard Smith and N. C. Stenseth, Evolution 38, 870 (1984). The ecological
time scale involves changes in population sizes and local extinction and
immigration, but a conserved number of species. The evolutionary time scale
deals with extinction and speciation events.

\item
S. Ellner and P. Turchin, Am. Nat. 145, 343 (1995)

\item
T. H. Keitt, and P. S. Marquet, J. Theor. Biol. 182, 161 (1996)

\item
T. H. Keitt and H. E. Stanley, Nature 393, 257 (1998)

\item
E. C. Pielou, {\em An Introduction to Mathematical Ecology}.
Wiley, 1969; Some recent, extensive field data analysis from marine
ecosystems has shown that this communities follow very well-defined
power law in species abundances with $N_s(n) \approx n^{-\gamma}$
with $\gamma \in (1,1.25)$ spanning three to four decades (S. Pueyo, 
unpublished data)

\item
R. V. Sol\'e and D. Alonso, Adv. Complex Systems 1, 203 (1998)

\item
Immigration is the main external source of perturbation in ecosystems not 
influenced by strong climate fluctuations.

\item 
N. G. Van Kampen, {\em Stochastic Processes in Physics and Chemistry}, 
Elsevier, 1981; C. W. Gardiner, {\em Handbook of Stochastic Methods} 
(2nd edition). Springer, Berlin (1990)

\item
M. Abramowitz and A. Stegun (eds), {\em Handbook of Mathematical Functions}, 
Dover, New York 1965.

\item 
In various intermediate expressions we have assumed that $\nu^{*}$ is not
an integer, but this final result is well defined for all meaningful ranges
of the parameters.

\item
Since $P_s(0) = (S-1)/S$, for $\lambda^*$ approaching zero, (12) should be 
approximated by $P_s(0) = ((S -1)/S) (\mu^* S)^{\lambda^*}$. Then, instead of 
(12), a correction is needed when $\lambda^*$ is tending to zero: 
$<S> - 1 = {\cal A} [C^*]^{-1+\epsilon(\mu)}$. As $\mu$ is tending to zero, 
$<S> = 1$, for high connectivities. 

\item
In (12), $C^*$ appears instead of the fraction of non-zero elements of 
the connectivity matrix, $C_{\pi}$. Both connectivities have exactly the same
meaning, i.e. the probability of having two species connected, as long as 
completely asymmetrical interaction between species is forbidden. If 
$\Omega_{ij}$ (the effect of $j$ on $i$ species) is non-zero, then 
$\Omega_{ji}$ may be extremely low but different from zero. The former 
assumption is very realistic in natural communities.

\end{enumerate}

\end{document}